\documentclass[11pt,a4paper,english,showpacs]{revtex4}

\usepackage{graphicx}
\usepackage{amsmath}
\usepackage{amssymb}
\usepackage{babel}
\usepackage{hyperref}



\begin{document}

\title{The Supersymmetric ($2+1$)D Noncommutative $CP^{(N-1)}$ Model in the Fundamental Representation}

\author{A. F. Ferrari, A. C. Lehum, A. J. da Silva, and F. Teixeira}

\affiliation{Instituto de F\'{\i}sica, Universidade de S\~{a}o Paulo\\
 Caixa Postal 66318, 05315-970, S\~{a}o Paulo - SP, Brazil}

\email{alysson, lehum, ajsilva, ftfilho@fma.if.usp.br}

\begin{abstract}
In this paper we study the noncommutative supersymmetric $CP^{(N-1)}$ model in $2+1$ dimensions, where the basic field is in the fundamental representation which, differently to the adjoint representation already studied in the literature, goes to the usual supersymmetric $CP^{(N-1)}$ model in the commutative limit. We analyze the phase structure of the model and calculate the leading and subleading corrections in a $1/N$ expansion. We prove that the theory is free of non-integrable UV/IR infrared singularities and is renormalizable in the leading order. The two-point vertex function of the basic field is also calculated and renormalized in an explicitly supersymmetric way up to the subleading order.
\end{abstract}


\maketitle

\section{Introduction}

The $CP^{(N-1)}$ model in $2+1$ dimensions was studied since the end of the 1970 decade, mainly because it is a reasonably simple scalar model which possesses gauge invariance, and it was found to reproduce several effects typical of more complicated gauge models in four spacetime dimensions, such as instantons solutions and confinement~\cite{D'Adda:1978uc,D'Adda:1978kp}. The crucial simplifying aspect of the $CP^{(N-1)}$ model is that the gauge field is non-dynamical at the classical level, its dynamics being entirely generated by quantum corrections. The possibility of studying in a simpler setting some of the most crucial aspects of gauge theories has been one of the main sources of interest in the model.

The phase structure of the $CP^{(N-1)}$ model, for example, was studied in~\cite{Arefeva:1980ms}, unveiling the existence of two phases. In one of them the symmetry $SU(N)$ is broken down to $SU(N-1)$, whereas in the other the model remains $SU(N)$ symmetric and mass generation occurs for the fundamental bosonic fields. Afterwards, it was found~\cite{Abdalla:1990qf} that the coupling to fermions preserves the two phases structure but the long range force is hindered by the fermionic fields. More recently, the supersymmetric $CP^{(N-1)}$ model was studied both in the Wess-Zumino gauge~\cite{Inami:2000eb} as well as using a manifestly supersymmetric covariant formulation~\cite{Cho:2003qn}.

On the other hand, in the past few years there have been a great deal of interest in quantum field theories defined over a noncommutative spacetime~\cite{Szabo:2001kg}. Sources of this interest are, among others, their relations with string theory~\cite{Seiberg:1999vs} and quantum gravity~\cite{Doplicher:1994tu}. In particular, gauge theories defined in a noncommutative spacetime have been intensively studied, and several interesting effects were found, such as UV/IR mixing~\cite{Minwalla:1999px,Hayakawa:1999yt} and strong restrictions of gauge groups and couplings to the matter~\cite{Matsubara:2000gr,Chaichian,Ferrari2}. The gauge invariance of the quantum corrections to the effective action also becomes very non-trivial in the noncommutative setting~\cite{Liu:2000ad,Pernici:2000va}.

Noncommutative gauge theories present a rich spectrum of phenomena, yet they can be also very complicated to deal with. So it was natural to investigate a simpler model with gauge invariance, such as the $CP^{(N-1)}$ model. When extending this model to the noncommutative spacetime, one finds more than one possibility of coupling the Lagrange multiplier and the gauge fields to the basic bosonic fields. In the non-supersymmetric case, the so-called fundamental and adjoint representations were considered in~\cite{Asano:2003ix}, using an $1/N$ expansion. In the fundamental representation, the model turned out to be renormalizable and free of dangerous infrared UV/IR singularities. However, in the adjoint representation, the appearance of non-integrable UV/IR divergences presented itself as an obstacle to the consistency of the model in higher orders of the $1/N$ expansion.

It is now well known that supersymmetry helps in avoiding the UV/IR problem in noncommutative field theories~\cite{Girotti:2000gc,Chepelev:2000hm,Matusis:2000jf}. Indeed, a supersymmetric extension of the noncommutative $CP^{(N-1)}$ model in the adjoint representation was studied in~\cite{Asano:2004vq}, showing that the aforementioned problem is not present. However, in that case, when the noncommutativity of the space is withdrawn the model does not return to the commutative supersymmetric $CP^{(N-1)}$, but to a free scalar theory, where the basic scalar field does not satisfy a constraint nor is coupled to an auxiliary gauge field. 

In the present work we are going to analyze the supersymmetric noncommutative $CP^{(N-1)}$ model in the fundamental representation, i.e., the case where the commutative limit really goes to the usual commutative supersymmetric $CP^{(N-1)}$ model. We shall study the phase structure of the model and the issue of the UV/IR mixing in the Feynman integrals, aiming at establishing its renormalizability at the leading order in the $1/N$ expansion. We will also look at the subleading corrections to the two-point vertex function of the scalar superfield, and show how the use of an explicitly supersymmetric quantization scheme avoids some difficulties with the usual (component) approach. 

Other aspects of the noncommutative supersymmetric $CP^{(N-1)}$ model in the fundamental representation were also focused in the recent literature. The structure of BPS and non-BPS solitons was first found to be quite similar to the commutative model~\cite{Lee:2000ey,Furuta:2002ty,Furuta:2002nv,Foda:2002nt}, but then novel solutions with no commutative counterparts were found~\cite{Otsu:2003fq}. This raised questions about the equivalence between the $CP^{1}$ and the nonlinear sigma model in the noncommutative case~\cite{Otsu:2004fz}. Alternatives to the above investigations, employing the Seiberg-Witten map, can be found in~\cite{Ghosh:2003ka} and~\cite{Govindarajan:2004kk}. More recently, the dynamics of the $CP^{(N-1)}$ model in a non-anticommutative spacetime has also been investigated~\cite{Inami:2004sq,Araki:2005nn}. 

This paper is organized as follows. In Section~\ref{sec1}, we present the model in superfield formulation and study its phase structure. Leading order corrections to the effective action of the auxiliary and gauge superfields are calculated in Section~\ref{sec3}, and the renormalizability of the model at the leading order is discussed in Section~\ref{sec4}. In Section~\ref{secphiphisub}, the quadratic effective action of the scalar superfield is discussed at the subleading order. Finally, Section~\ref{conc} contains our conclusions and final remarks. In the Appendix, we explicitly give the component formulation of the superfield model studied by us.

\section{The Noncommutative Supersymmetric $CP^{(N-1)}$ Model}\label{sec1}

The bosonic $CP^{(N-1)}$ model, in commutative spacetime, when the matter field is in the fundamental representation, is defined by the action~\cite{Arefeva:1980ms}
\begin{eqnarray}\label{c3eq1}
S=\int~d^3x\Big{\{}\overline{ D^{\mu} \phi}_a D_{\mu}\phi_a+\sigma\Big(\phi_a\bar\phi_a-
{\frac{N}{g_0}}\Big)\Big{\}}~,
\end{eqnarray}

\noindent
where $\phi_a$ is an N-uple of scalar fields and $\sigma$ is a scalar Lagrange multiplier which 
enforces the constraint $\phi\bar\phi={N/g_0}$. The covariant derivative is $D^{\mu}=\partial^{\mu}-iA^{\mu}$, $A_{\mu}$ being an auxiliary vector gauge field, which classically is the composite field 
$A^{\mu}=i\phi \overset{\leftrightarrow}{\partial^{\mu}} \bar\phi/2\phi\bar\phi$.
The spacetime index $\mu$ runs over $0, 1, 2$, and we use the metric $g^{\mu\nu}=(-,+,+)$.

The model defined by~(\ref{c3eq1}) can be generalized to a noncommutative spacetime where coordinates satisfy \begin{equation}
[x^{\mu},x^{\nu}]=i\Theta^{\mu\nu}
\end{equation}

\noindent
by substituting into~(\ref{c3eq1}) the usual product of functions by the $*$-Moyal product~\cite{moyal}. The divergence structure of the noncommutative $CP^{(N-1)}$ model has been extensively discussed in~\cite{Asano:2003ix,Asano:2004vq}. Here we shall focus on a noncommutative supersymmetric extension 
of the model~(\ref{c3eq1}) which, adopting the conventions of~\cite{Gates:1983nr}, is given by
\begin{eqnarray}\label{c3eq2}
S  =  -\int~d^5z\Big{\{}~{\frac{1}{2}}~\overline{\nabla^{\alpha}\Phi_a}*\nabla_{\alpha}\Phi_a~+~
  \Sigma *\Big(\Phi_a*\bar\Phi_a-{\frac{N}{g_0}}\Big)\Big{\}}\,.
\end{eqnarray}

\noindent
In this expression, $z=(x^{\mu},\theta^{\alpha})$, where $x^{\mu}$ with $\mu=0,1,2$ and $\theta^{\alpha}$ with 
$\alpha=1,2$ are, respectively, the bosonic and the grassmanian superspace coordinates, $\Phi_a$ is an N-uple of scalar superfields, $\Sigma$ is a scalar Lagrange multiplier superfield, and $A_{\alpha}$ is a two-component 
spinor auxiliary gauge superfield. The supercovariant spinorial derivative is given by $\nabla_{\alpha}
=D_{\alpha}-iA_{\alpha}$, where $D_{\alpha}=\partial_{\alpha}+i\theta^{\beta}\partial_{\beta\alpha}$. We remark that the $*$-Moyal product in~(\ref{c3eq2}) is defined by~\cite{Chu:1999ij}
\begin{equation}\label{c3eq2a}
f(x,\theta)*g(x,\theta)\,=\,\exp\Big (\frac{i}{2}\Theta^{\mu\nu}\frac{\partial}
{\partial x^{\mu}}
\frac{\partial}{\partial y^{\nu}}\Big)f(x,\theta)g(y,\theta)|_{x=y}\,,
\end{equation}

\noindent
affecting only the bosonic coordinates $x^{\mu}$, which are noncommutative, whereas the grassmanian coordinates satisfy the usual anticommutativity rule $\{\theta^a,\theta^b \}=0$~\footnote{Notice, however, that one can consider the extension of the noncommutativity to the grassmanian superspace coordinates in four spacetime dimensions~\cite{Seiberg}. Recently, this idea has also been extended to three-dimensional spacetime~\cite{ournac}}. To avoid troubles with unitarity, we shall restrict the 
noncommutativity to the spatial bosonic coordinates, what amounts to consider $\Theta^{0i}=0$~\cite{Gomis:2000zz} \footnote{It is interesting to remember that unitarity violations are a peculiar aspect of the approach for noncommutative field theories we are considering. There are alternative proposals which do not suffer from this issue, see~\cite{Doplicher:1994tu,Bahns:2002vm,Liao:2002xc,Balachandran:2004rq}}. 

The action above is U(N) globally invariant and U(1) gauge invariant. The 
infinitesimal gauge transformations are given by,
\begin{eqnarray}\label{c3eq2b}
\Phi_a&\rightarrow&\Phi^{\prime}_a=(1+iK)*\Phi_a~,\nonumber\\
\bar\Phi_a&\rightarrow&\bar\Phi^{\prime}_a=
\bar\Phi_a*(1-iK)~,\\
A_{\alpha}&\rightarrow& A^{\prime}_{\alpha}=
A_{\alpha}+D_{\alpha}K+i[K,A_{\alpha}]_{*}~,\nonumber\\
\Sigma&\rightarrow&\Sigma^{\prime}=\Sigma+i[K,\Sigma]_{*},\nonumber
\end{eqnarray}

\noindent
where $K$ is a real scalar superfield. The ordering of $\Phi$ and $\bar\Phi$ in the constraint term in 
Eq.~(\ref{c3eq2}) implies in the need of $\Sigma$ being transformed under the gauge transformation 
in order to maintain the invariance of the action (observe that the 
transformation of $\Sigma$ disappears when we take the commutative limit 
$\Theta\rightarrow0$, as it should). Had we chosen the opposite order 
for $\Phi$ and $\bar\Phi$, then $\Sigma$ would not need to transform, but a 
mixing between the fields $\Sigma$ and $A^{a}$ would appear already in 
leading $1/N$ approximation (we will return to this point later).

By writting the original (unrenormalized) superfields in terms of 
renormalized ones through the definitions, $\Phi=Z^{1/2}_1\Phi_R$, $A^{\alpha}=
Z_3A^{\alpha}_R$, and $\Sigma=Z_2\Sigma_R$, the action gets written as,
\begin{eqnarray}\label{c3eq17d}
S & = & -\,\int d^5z \Big\{ \frac{Z_1}{2}D^{\alpha}\bar\Phi_{Ra}
D_{\alpha}\Phi_{Ra}+Z_2\Sigma\Big(Z_1\Phi_{Ra}\bar\Phi_{Ra}-\frac{N}{g_0}
\Big)\nonumber\\
& - & \frac{iZ_1Z_3}{2}\Big(D_{\alpha}\bar\Phi_{Ra}A_{R\alpha}\Phi_{Ra}
+\bar\Phi_{Ra}A_{Ra}D^{\alpha}\Phi_{Ra}\Big)+\frac{1}{2}Z_1Z^2_3\bar\Phi_{Ra}
A^2_R\Phi_{Ra}\Big\}~.
\end{eqnarray}

\noindent
In this equation and in the remaining of this paper, we will not explicitly indicate the $*$-Moyal product, which should be understood to be present when multiplying fields in configuration space.
Wave functions and coupling constant counterterms are defined through
\begin{eqnarray}\label{c3eq17e}
Z_1&=&1+\delta_{\Phi}~,\nonumber\\
Z_1Z_3&=&(1+\delta_{\Phi})(1+\delta_A)=1+\delta_e~,\nonumber\\
Z_1Z^2_3&=&(1+\delta_{\Phi})(1+\delta_A)^2=1+\delta_b~,\nonumber\\
Z_1Z_2&=&(1+\delta_{\Phi})(1+\delta_{\Sigma})=1+\delta_c~,\nonumber\\
Z_2/g_o&=&\mu/g+\delta_{g}~,
\end{eqnarray}

\noindent
where $\mu$ is an arbitrary parameter with dimension of mass and $g$ is the 
adimensional renormalized coupling constant. Substituting these expressions in Eq.~(\ref{c3eq17d}) and omitting the subindex R that indicates renormalized fields, we have
\begin{eqnarray}\label{c3eq17f}
S& = &-\int d^5z \Big\{ \frac{1}{2}D^{\alpha}\bar\Phi_a D_{\alpha}\Phi_a
+\Sigma\Big(\Phi_a\bar\Phi_a-\frac{N \mu}{g}\Big)
\nonumber\\
& + &\frac{\delta_{\Phi}}{2}D^{\alpha}\bar\Phi_aD_{\alpha}\Phi_a
-i\frac{\delta_e}{2}(D^{\alpha}\bar\Phi_aA_{\alpha}\Phi_a
+\bar\Phi_aA_{\alpha}D^{\alpha}\Phi_a)
\nonumber\\
& + &\frac{\delta_b}{2}\bar\Phi_aA^{\alpha}A_{\alpha}\Phi_a
-N\delta_g\Sigma+\delta_c\Sigma\Phi_a\bar\Phi_a\Big\} 
\end{eqnarray}

\noindent
From now on, we will not explicitly write the R subindex; all fields will be understood to be the renormalized ones. 

To study the phase structure of the model let us suppose that
the superfields $\Sigma$ and $\Phi$ acquire constant non-vanishing vaccum expectation values (VEVs) 
$\langle\Sigma(x,\theta)\rangle=m$ and $\langle\Phi(x,\theta)\rangle=\sqrt{N}v$, 
this last one, for simplicity, supposed to be in the $a=N$ component~\footnote{We shall not consider possible quantum phases that could arise from classical noncommutative solitonic solutions, which can be singular at small $\Theta$~\cite{Gopakumar:2000zd}. }. These VEVs will play the role of order parameters identifying the different phases we will find. By redefining the fields in term of new fields that have zero VEVs,
\begin{eqnarray}\label{c3eq8}
\Phi_a(x,\theta)& \longrightarrow & \Phi_a(x,\theta)\,,\,\,a=1,\ldots,N-1,\nonumber\\
\Phi_N(x,\theta)& \longrightarrow & \Phi_N(x,\theta)+{v \sqrt{N}}~,\nonumber\\
\bar\Phi_N(x,\theta)& \longrightarrow & \bar\Phi_N(x,\theta)+{\bar{v}
\sqrt{N}}~,\\
\Sigma(x,\theta)& \longrightarrow & \Sigma(x,\theta)+m~,\nonumber\\
A_\alpha(x,\theta)& \longrightarrow & A_\alpha(x,\theta),\nonumber
\end{eqnarray}

\noindent
the action of Eq.~(\ref{c3eq17f}) is written as 
\begin{eqnarray}\label{c3eq9}
S & = & \int~d^5z\Big\{~\bar\Phi_a(D^2-m)\Phi_a~-~\Sigma(\Phi_a\bar\Phi_a-\frac {N \mu}{g}
+N~v\bar{v})-{1\over 2}\bar\Phi_aA^{\alpha}A_{\alpha}\Phi_a
\nonumber\\
& + & {i \over 2}\big[D^{\alpha}\bar\Phi_aA_{\alpha}\Phi_a+\bar\Phi_a A_{\alpha}D^{\alpha}
\Phi_a+v\sqrt{N}D^{\alpha}\bar\Phi_NA_{\alpha}-\bar{v}\sqrt{N}A^{\alpha}
D_{\alpha}\Phi_N\big]
\\
& - & {\sqrt{N}\over 2}\bar{v}A^{\alpha}A_{\alpha}\Phi_N+{\sqrt{N}\over
  2}v\bar\Phi_NA^{\alpha}A_{\alpha}+ {N\over 2}\bar{v}vA^{\alpha}A_{\alpha}
-m\sqrt{N}(v\bar\Phi_N+\bar{v}\Phi_N)
\nonumber\\
 & - &  \Sigma \Phi_N \bar v \sqrt{N} +  \Sigma \bar \Phi_N v \sqrt{N} \Big\}~+~S_{CT},\nonumber
\end{eqnarray}

\noindent
where $S_{CT}$ is a short for the counterterms action, and $D^2 \equiv \frac{1}{2}D^\alpha D_\alpha$. The propagator for the first 
$(N-1)$ components of $\Phi_a$ is given by
\begin{equation}\label{c3eq9a}
\langle T \, \Phi_a(p,\theta_1)\bar\Phi_b(-p,\theta_2)\rangle=
-i \delta_{ab} \left( \frac{D^2+m}{p^2+m^2} \right) \delta_{12}\,,
\end{equation}

\noindent
where $\delta_{12}\equiv\delta^2(\theta_1-\theta_2)$. The interaction vertices of the theory are
\begin{eqnarray}
\Sigma\Phi\bar\Phi\,\, &\longrightarrow& -ie^{-{i}k_2\wedge k_3}
\,\,\Sigma(1)\Phi(2)\bar\Phi(3)\,,\nonumber\\
\bar\Phi A^{\alpha}D_{\alpha}\Phi+\cdots\,\,&\longrightarrow&
\frac{1}{2}e^{-{i}k_2\wedge k_3}
A^{\alpha}(1)\,D_{\alpha}\left[\Phi(2)\bar\Phi(3)\right]\,,\nonumber\\
\bar\Phi A^{\alpha}A_{\alpha}\Phi\,\,&\longrightarrow&
\frac{i}{2} \cos(k_1\wedge k_2)e^{-{i}k_3\wedge k_4}\,
A^{\alpha}(1)A_{\alpha}(2)\Phi(3)\bar\Phi(4)\,,
\end{eqnarray}

\noindent
where $k\wedge p \equiv \frac{1}{2} k_{\mu}\Theta^{\mu\nu}p_{\nu}$ and $A^{\alpha}(1)$ denotes $A^{\alpha}(k_1,\theta)$, and similarly for the other fields. 

\begin{figure}[ht]
\begin{center}\includegraphics[width=0.6\textwidth]{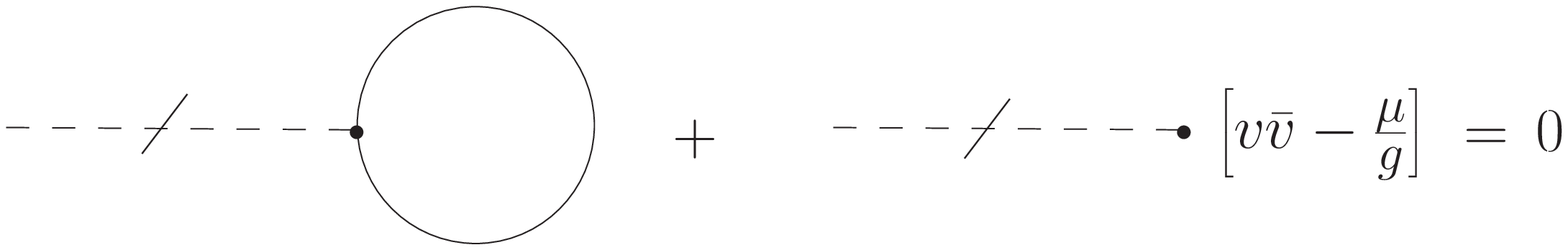}
\end{center}
\caption{Gap equation for the $\Sigma$ superfield, including the tadpole loop correction. In this and in the following figures, solid lines represent $\Phi_a$ propagators. Dashed cut lines represent the $\Sigma$ field.} \label{figgap}
\end{figure}

The condition that, in leading order of $1/N$, the redefined fields 
$\Sigma$ and $\Phi$ have zero vacuum expectation values imply in the equations, 
\begin{eqnarray}\label{c3eq10}
&&i\int_{\epsilon}~{d^3k\over(2\pi)^3}{1\over k^2+m^2}
+\frac{\mu}{g}+\delta_g(\epsilon)-v\bar{v}=0~,\nonumber\\[0.5cm]
&&m\bar{v}=mv=0~.
\end{eqnarray}

\noindent
where the $\epsilon$ in the integral symbol and in $\delta_g$ represent
an ultraviolet regulator. The first gap equation is represented graphically in Fig.~\ref{figgap}. 
These equations are the same as the  
corresponding ones for the supersymmetric commutative model~\cite{Inami:2000eb}. The dependence on $\Theta$ of the underlying noncommutativity of the spacetime, manifested through the phase factors appearing in the vertices, disappear due to the vanishing of the momentum entering through the external leg of $\Sigma$ or $\Phi_N$. In particular, this fact ensures that UV/IR infrared singularities do not appear in the gap equation. The scalar integral in (\ref{c3eq10}) can be performed using dimensional reduction~\cite{Siegel:1979wq}, leading to
\begin{equation}
\label{gapint}
i\,\int_{\epsilon}~{d^3k\over(2\pi)^3}{1\over k^2+m^2} \,=\, \frac{ | m |}{4\pi}
\end{equation}

\noindent
hence the counterterm $\delta_g$ turns out to be finite, providing an arbitrary finite renormalization of the gap equation, which now reads,
\begin{eqnarray}\label{c3eq11}
\frac{|m|}{4\pi} +\frac{\mu}{g}+\delta_g
-v\bar{v}=0~,\nonumber\\[0.5cm]
m\bar{v}=mv=0~.
\end{eqnarray}

One convenient choice for the counterterm is $\delta_g=-{\mu}/{4\pi}$, where $\mu > 0$ is the same mass scale introduced in~(\ref{c3eq17e}), so that the solution of the first of Eqs.~(\ref{c3eq11}) can be written as
\begin{equation}
g=\frac{\mu}{\bar{v} v+\frac{\mu-|m|}{4\pi}}\,. 
\end{equation}

\noindent
From the second of Eqs.~(\ref{c3eq11}) we see that the model presents two phases,
\begin{enumerate}
 \item A broken U(N) phase in which $\Phi_N$ has a vacuum expectation value $\langle\Phi_N\rangle=N^{1/2}v\neq0$ and the fields $\Phi$ remain massless ($m=0$). This happens for 
\begin{equation}
g= \frac{4\pi}{(1+4\pi\bar{v}v/\mu)}<4\pi\,. 
\end{equation}
 \item A symmetric phase in which $\Phi$ has an induced mass $m\neq 0$ but $\langle\Phi\rangle=0$. This happens for
\begin{equation}
g=\frac{4\pi}{1-|m/\mu|} >4\pi\,.\label{c3eq12}
\end{equation}
\end{enumerate}

From Eq.~(\ref{c3eq12}) one can immediately calculate the $\beta$ function in the symmetric phase,
\begin{equation}
\beta(g)\equiv \mu \frac{dg}{d\mu}=4\pi \frac{|m/\mu|}{(1-|m/\mu|)^2}
=g\left(1-\frac{g}{4\pi}\right).\label{c3eq13}
\end{equation}

\noindent
As can be read from this formula, $\beta$ goes to zero for $\mu \rightarrow 
\infty$, characterizing an ultraviolet fixed point at $g=4\pi$. This result is the same as the one for 
the corresponding commutative model~\cite{Inami:2000eb}. The same analysis for the behavior of the $\beta$ function in the broken phase leads to similar conclusions.

We stress that the choice $\delta_g=-{\mu}/{4\pi}$ is convenient, but not essential. Any value of $\delta_g$ provided that $\delta_g \, < \, 0$ leads to the same phase structure, only the value of the critical $g$ changes. If $\delta_g \, > \, 0$, on the other hand, the symmetric phase does not exist, while for $\delta_g \, = \, 0$, the model does not have the broken phase. Finally, one could choose other regularization to calculate the divergent integral in Eq.~(\ref{gapint}), in which case it could be necessary the counterterm $\delta_g$ to contain an infinite renormalization to render the gap equation~(\ref{c3eq10}) finite. Even in this case, the above considerations would apply without any changes.

\section{Effective propagators at leading order}
\label{sec3}

As we are mainly interested in studying the renormalization of the model
and the two phases have the same ultraviolet behavior~\cite{coleman} 
we will work in the symmetric phase from now on, so that the action reads
\begin{eqnarray}\label{actionp}
S & = & \int~d^5z\Big\{~\bar\Phi_a(D^2-m)\Phi_a~-~\Sigma\left(\Phi_a\bar\Phi_a-\frac {N \mu}{g}\right)-\frac{1}{2}\bar\Phi_aA^{\alpha}A_{\alpha}\Phi_a\nonumber\\
  & + & {i \over 2}\big[D^{\alpha}\bar\Phi_aA_{\alpha}\Phi_a+\bar\Phi_a A_{\alpha}D^{\alpha}
\Phi_a\big]\Big\}~+~S_{CT},
\end{eqnarray}

\noindent
where $g$ is related to $m$ and $\mu$ by Eq.~(\ref{c3eq12}), and
\begin{eqnarray}\label{sct}
S_{CT}& = & \int~d^5z\Big\{ {\delta_{\Phi}}\bar\Phi_aD^2\Phi_a-m\delta_c \bar\Phi_a\Phi_a
-i\frac{\delta_e}{2}(D^{\alpha}\bar\Phi_aA_{\alpha}\Phi_a
+\bar\Phi_aA_{\alpha}D^{\alpha}\Phi_a)
\nonumber\\
& - &\frac{\delta_b}{2}\bar\Phi_aA^{\alpha}A_{\alpha}\Phi_a
+N\delta_g\Sigma
-\delta_c\Sigma\Phi_a\bar\Phi_a\Big\}\,.
\end{eqnarray}

From these equations, we see that the propagator of the $\Phi_N$ is also given by Eq.~(\ref{c3eq9a}). This fact simplifies the perturbative calculations in the next subsections. 

\subsection{The two point effective action of the  $\Sigma$ Field}
\label{tpsigma}

\begin{figure}[ht]
\begin{center}\includegraphics[width=0.3\textwidth]{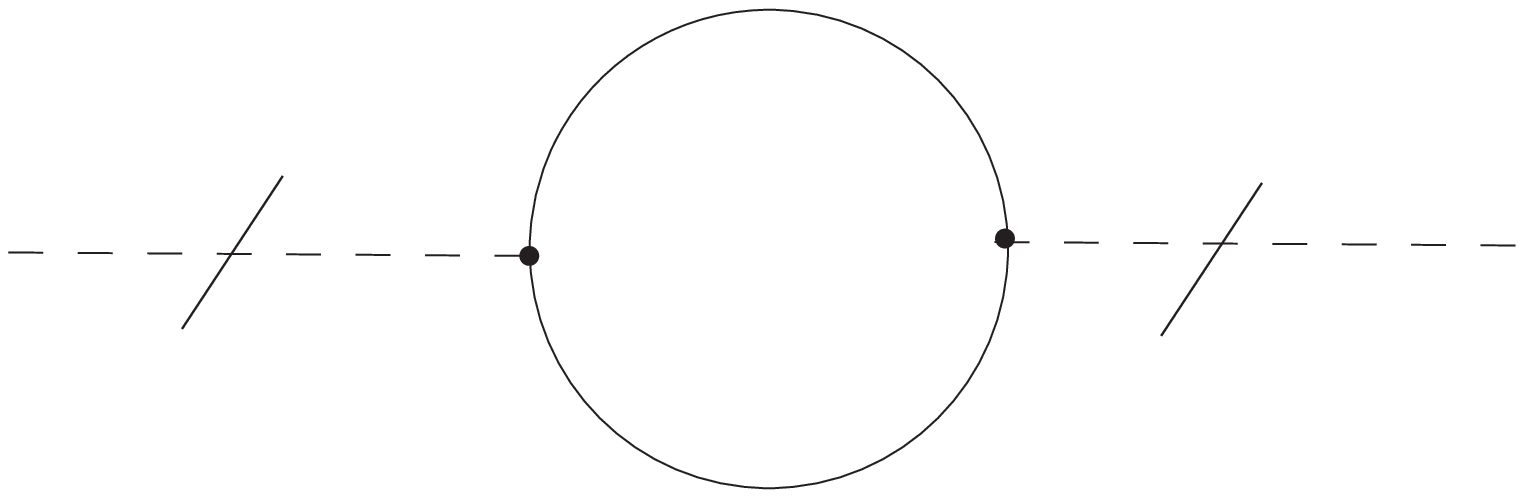}
\end{center}
\caption{Leading order of $1/N$ contribution to the effective propagator of the $\Sigma$ field.} \label{Fig1}
\end{figure}

As we read from the Eq.~(\ref{actionp}), classically the $\Sigma$ field is 
purely a constraint field without dynamics. However, in leading order of
$1/N$ it acquires a (nonlocal) kinetic term becoming a propagating 
field. In this approximation, the radiative corrections to its two point 
effective action (see Fig.~\ref{Fig1}) is given by
\begin{eqnarray}\label{c3eq20}
\Gamma^{(2)}_{\Sigma}={1 \over 2}\int{d^3p\over(2\pi)^3}d^2\theta~\Sigma(-p,\theta)\left[Nf(p)(D^2+2m)\right]
\Sigma(p,\theta)~.
\end{eqnarray}

\noindent
The exponential factors in the two vertices cancel between themselves and the result is similar to the commutative one. The nonlocal character of this action is explicit in the factor $f(p)$, 
\begin{eqnarray}\label{c3eq21}
f(p)&=&\int{d^3k\over(2\pi)^3}{1\over[(k+p)^2+m^2](k^2+m^2)}\nonumber\\
    &=&-{1\over 4\pi\sqrt{p^2}}\,{\rm arctg}\left({1\over 2}\sqrt{p^2\over
  m^2}\right)=\left\{ \begin{array}{ll}
-1/ 8|p|\,\, {\rm for }\,\,\,\, p\rightarrow~\infty \\
-1/ 8 \pi m \,\, {\rm for }\,\,\,\, p\rightarrow~0 
\end{array} \right.\,.
\end{eqnarray}

\noindent
The finiteness of $\Gamma^{(2)}_{\Sigma}$ is consistent with the absence of a $\Sigma^2$ counterterm in the original action. From Eq.~(\ref{c3eq20}), we arrive at the following $\Sigma$ propagator,
\begin{eqnarray}\label{c3eq22}
\langle T\,\Sigma(p,\theta_1)\Sigma(-p,\theta_2) \rangle={i\over
  N}{(D^2-2m)\over f(p)(p^2+4m^2)}\delta_{12}~,
\end{eqnarray}

\noindent
which is regular in the infrared ($p^2 \rightarrow 0$) while decreasing as $1/p$ in the ultraviolet limit ($p^2 \rightarrow \infty$).

\begin{figure}[ht]
\begin{center}\includegraphics[width=0.4\textwidth]{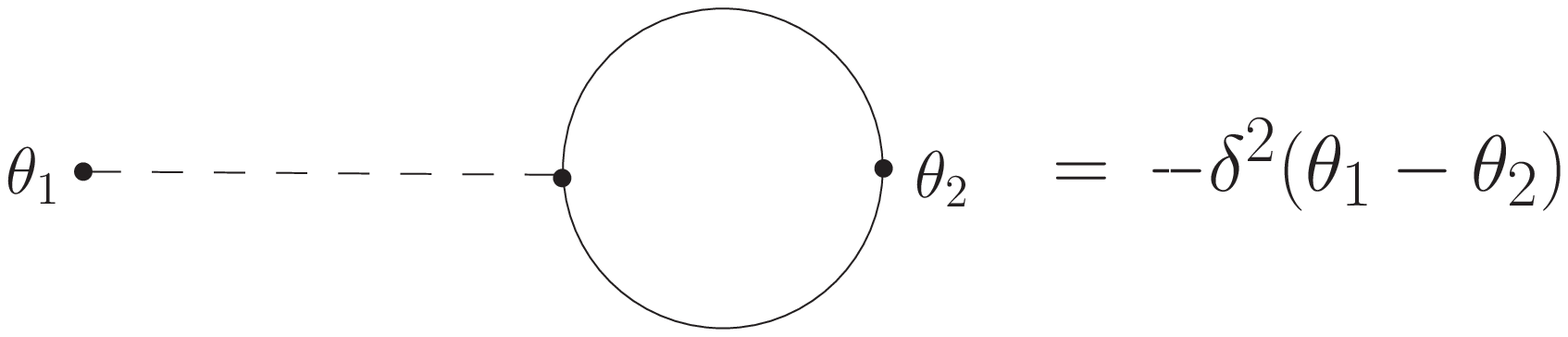}
\end{center}
\caption{This supergraph identity, discussed in~\cite{Arefeva:1980ms} in the non-supersymmetric $CP^{(N-1)}$ model, secures the cancellation of several divergences when the perturbation theory is expanded to include one particle \textit{reducible} graphs. Without this identity, the proof of renormalizability of the $CP^{(N-1)}$ model at higher orders becomes unfeasible.} \label{figarefeva}
\end{figure}

Since, by definition, the $\Sigma$ effective propagator is \textit{minus} the inverse of the kernel in Eq.~(\ref{c3eq20}), in the supergraph formalism we still have the identity represented graphically in Fig.~\ref{figarefeva}, which is known to hold in the usual commutative $CP^{(N-1)}$ model~\cite{Arefeva:1980ms}. This powerfull identity is very important in the study of subleading quantum corrections to the vertex functions of the model, as we will comment in Section~\ref{secphiphisub}.

\subsection{The two point effective action of the spinorial gauge potential}\label{tpA}

\begin{figure}[ht]
\begin{center}\includegraphics[width=0.5\textwidth]{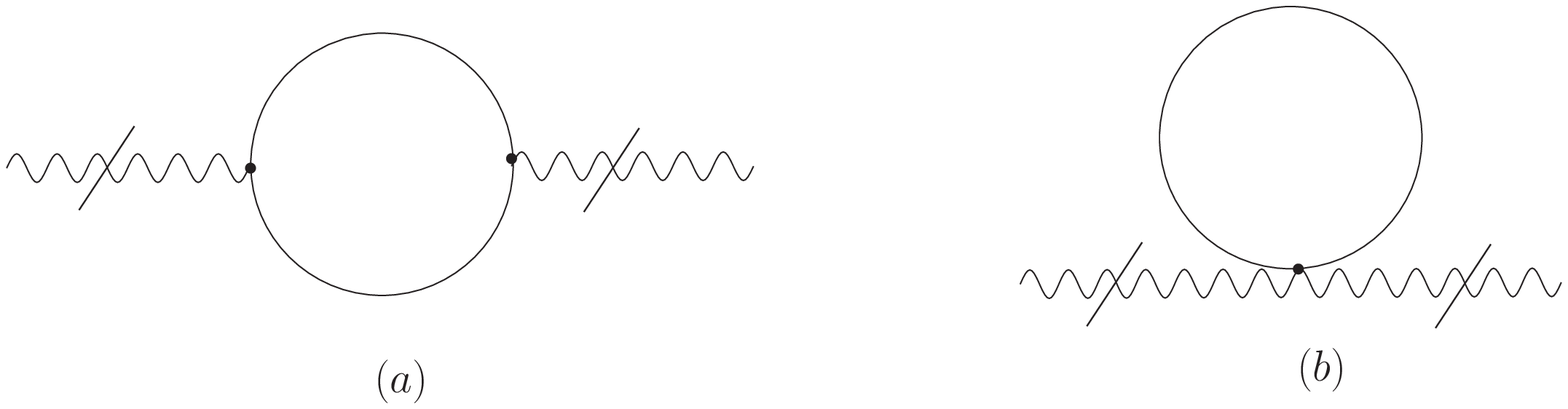}
\end{center}
\caption{Leading order of $1/N$ contribution to the spinorial superfield effective propagator. Wavy cut lines represent the $A$ field.} \label{Fig2}
\end{figure}

Another field whose dynamics is generated only at the quantum level is the spinorial superfield $A_{\alpha}$.
The leading $1/N$ radiative correction to its two point effective action is represented in Fig.~\ref{Fig2}. 
The contribution of the graph~\ref{Fig2}a gives
\begin{equation}\label{c3eq23c}
\Gamma^{(2)}_{\ref{Fig2}a} = {N\over 2}\int{d^3p\over(2\pi)^3}d^2\theta
\int_\epsilon{d^3k\over(2\pi)^3}\frac{C^{\alpha\beta} }{k^2+m^2}\, ~A_{\alpha}(p,\theta)A_{\beta}(-p,\theta)~,
\end{equation}

\noindent
while the graph~\ref{Fig2}b yields
\begin{eqnarray}\label{c3eq24}
\Gamma^{(2)}_{\ref{Fig2}b} & = & -{N\over 2}\int{d^3p\over(2\pi)^3}d^2\theta~\int_\epsilon{d^3k
\over(2\pi)^3}{1\over[(k+p)^2+m^2](k^2+m^2)}\nonumber\\
&\times&\Big[(k^2+m^2)C^{\alpha\beta}+(k^{\alpha\beta}+mC^{\alpha\beta})D^2+
{1\over 2}(k^{\gamma\beta}+mC^{\gamma\beta})D_{\gamma}D^{\alpha}\Big]\nonumber\\
&&\times A_{\alpha}(p,\theta)A_{\beta}(-p,\theta)~.
\end{eqnarray}

\noindent
Adding the two contributions above we get
\begin{eqnarray}\label{c3eq24a}
\Gamma^{(2)}_{\ref{Fig2}} & = & -{N\over 2}
\int{d^3p\over(2\pi)^3}d^2\theta~\int_\epsilon{d^3k\over(2\pi)^3}{1\over[(k+p)^2+m^2](k^2+m^2)}
\left(k^{\alpha\beta}+mC^{\alpha\beta}\right)\nonumber\\
&\times&\left(D^2A_{\alpha}(p,\theta)A_{\beta}(-p,\theta)+{1\over 2}D_{\alpha}D^{\gamma}A_{\gamma}(p,\theta)A_{\beta}(-p,\theta)\right)~.
\end{eqnarray}

\noindent
Individually, the graphs in Fig.~\ref{Fig2} are linearly ultraviolet divergent, but their leading divergences cancel between themselves, so that $\Gamma^{(2)}_{\ref{Fig2}}$ contains at most logarithmic divergences. In fact, it turns out to be finite, since $\int_\epsilon d^3k \, \frac{k^{\alpha \beta}}{\left(k^2\right)^2} = 0$ due to the symmetric integration in the loop momentum, and after using the identity
\begin{equation}\label{c3eq25}
\int_{\epsilon}{d^3k\over(2\pi)^3}{k^{\alpha\beta}\over[(k+p)^2+m^2](k^2+m^2)}= 
-\frac{p^{\alpha\beta}}{2}\int_{\epsilon}{d^3k\over(2\pi)^3}{1\over[(k+p)^2+m^2](k^2+m^2)}~,
\end{equation}

\noindent
Eq.~(\ref{c3eq24a}) can be written as 
\begin{eqnarray}\label{c3eq26}
\Gamma^{(2)}_{\ref{Fig2}} & = & \frac{N}{4} \int{d^3p\over(2\pi)^3}d^2\theta~ f(p)\left(-p^{\alpha\beta}+2mC^{\alpha\beta}\right)
A_{\beta}(p,\theta)W_{\alpha}(-p,\theta)~,
\end{eqnarray}

\noindent
where
\begin{eqnarray}\label{c3eq27}
W^{\alpha}={1\over 2}D^{\beta}D^{\alpha}A_{\beta}={1\over
  2}D^{\alpha}D^{\beta}A_{\beta}+D^2A^{\alpha}~,
\end{eqnarray}

\noindent
corresponds to the linear part of the (noncommutative) Maxwell superfield 
strength. Using the relation $D^2W_{\alpha}=-{p_{\alpha}}^{\beta}W_{\beta}$~\cite{Gates:1983nr}, we can write Eq.~(\ref{c3eq26}) as 
\begin{eqnarray}\label{c3eq30}
\Gamma^{(2)}_{3a+3b}  & = & {N\over 4}\int{d^3p\over
  (2\pi)^3}d^2\theta f(p)\big(W^{\alpha}W_{\alpha}+
  2mA^{\alpha}W_{\alpha}\big)~,
\end{eqnarray}

\noindent
where we omitted the explicit arguments of the external fields since they follow the same pattern as in the previous equations. In Eq.~(\ref{c3eq30}) the first and second terms are induced nonlocal Maxwell and 
Chern-Simons terms. The local limit obtained by the approximation $f(p)\simeq
f(0)=-1/8\pi m$ gives for the coefficient of the induced Chern-Simons term 
the value $N/16\pi$. 

To calculate the propagator of $A^{\alpha}$ we need to fix a gauge.  
One frequent choice in the literature is the Wess Zumino gauge, which amounts to 
choosing the components $\chi_{\alpha}$ and $B$ of the superpotential $A_{\alpha}$ 
(see Eq.~(\ref{c3eq3a}) in the Appendix) as vanishing; this choice greatly simplifies the calculation in terms of component fields, but it breaks manifest supersymmetry and can lead into difficulties. So, we will fix the gauge by adding to the action given by Eq.~(\ref{c3eq30}) a covariant nonlocal gauge fixing term,
\begin{eqnarray}\label{c3eq31}
S_{GF} & = & {N\over 8\xi}\int{d^3p\over
  (2\pi)^3}d^2\theta f(p)D^{\beta}A_{\beta}D^2D^{\alpha}A_{\alpha}~,
\end{eqnarray}

\noindent
and the corresponding Faddeev-Popov action,
\begin{eqnarray}\label{c3eq35}
S_{FP}=-{N\over 4}\int{d^3p\over
  (2\pi)^3}d^2\theta~f(p)\left(c^{\prime}D^2c-ic^{\prime}D^{\alpha}[A_{\alpha},c]\right)~.
\end{eqnarray}

\noindent
With this gauge choice, the part of the action quadratic in $A^{\alpha}$ turns out to be
\begin{equation}\label{c3eq32}
\Gamma^{(2)}_A =  -{N\over 8}\int{d^3p\over(2\pi)^3}d^2\theta~f(p) 
A_{\alpha}\left[D^{\beta}D^{\alpha}(D^2+2m)+{1\over \xi} D^{\alpha}
D^{\beta}D^2\right]A_{\beta}~,
\end{equation}

\noindent
from which follows the propagator
\begin{equation}\label{c3eq33}
\langle T~A^{\alpha}(p,\theta_1)A^{\beta}(-p,\theta_2)\rangle=
{i\over Nf(p)}\left[{(D^2-2m)D^{\beta}D^{\alpha}\over p^2(p^2+4m^2)}
+\xi{D^2D^{\alpha}D^{\beta}\over p^4}\right]\delta_{12}~,
\end{equation}

\noindent
or in another form, that will be usefull for following calculations,
\begin{eqnarray}\label{c3eq34}
&&\langle T~A^{\alpha}(p,\theta_1)A^{\beta}(-p,\theta_2) \rangle  = 
{i\over Nf(p)}\Big[-{2mp^{\alpha\beta} \over p^2(p^2+4m^2)}+\left(\frac{1}{p^2+4m^2}-{\xi\over p^2}\right)C^{\alpha\beta}
\nonumber\\
&&  +  {p^{\alpha\beta}\over p^2}\left(\frac{1}{p^2+4m^2}+{\xi\over p^2}\right)D^2+
{2mC^{\alpha\beta}\over p^2(p^2+4m^2)}D^2\Big]\delta_{12}.
\end{eqnarray}

\noindent
The ghost propagator obtained from the Faddeev-Popov action is 
\begin{eqnarray}\label{c3eq36}
\langle  T\, c^{\prime}(p,\theta_1)c(-p,\theta_2) \rangle=-i{4\pi\over
  N}{D^2\over f(p)p^2}\delta_{12}~,
\end{eqnarray}

\noindent
and its contribution only appears at the $1/N^2$ order, so that up to the order of approximation we are considering, its contribution will not appear.

\subsection{Is there an $A^{\alpha}\Sigma$ mixing?} \label{mixing}

\begin{figure}[ht]
\begin{center}\includegraphics[width=0.3\textwidth]{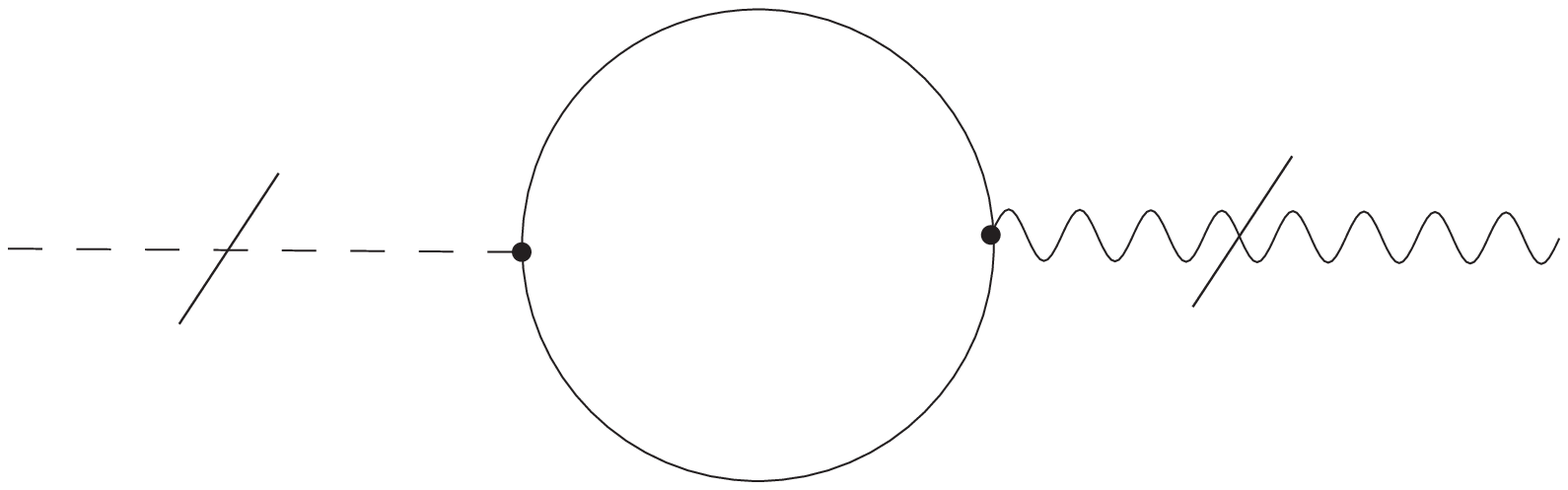}
\end{center}
\caption{Supergraph that could imply into a leading order process mixing the spinorial and the $\Sigma$ superfields.} \label{Fig3}
\end{figure}

From the action in Eq.~(\ref{actionp}), we see that an $1/N$ order process
mixing $A^{\alpha}$ and $\Sigma$ is in principle possible, what 
would result in a mixed propagator. The graph contributing to this process 
is represented in Fig.~\ref{Fig3} and the corresponding Feynman amplitude,
\begin{eqnarray}\label{c3eq37}
&&\Gamma^{(2)}_{A\Sigma} = \langle \, T:
-i\int d^5z~\Sigma(z)\Phi_a(z)\bar\Phi_a(z):\nonumber\\
&& \times  : {i\over 2!}\int d^5 z^{\prime}~ {i\over 2}\Big[D^{\alpha}\bar\Phi_b(z^{\prime})A_{\alpha}(z^{\prime})\Phi_b(z^{\prime})+\bar\Phi_b(z^{\prime})
A_{\alpha}(z^{\prime})D^{\alpha}\Phi_b(z^{\prime})\Big]: \, \rangle \,,
\end{eqnarray}

\noindent
can be separated in two terms, $\Gamma^{(2)}_{A\Sigma}=\Gamma^{(2)}_{A\Sigma,1}+\Gamma^{(2)}_{A\Sigma,2}$, the first one giving
\begin{equation}\label{c3eq38}
\Gamma^{(2)}_{A\Sigma,1}=i{N\over 4}\int{d^3p\over(2\pi)^3}d^2\theta \int_\epsilon{d^3k\over(2\pi)^3}
{(k^{\beta\alpha}-mC^{\beta\alpha})\over [(k+p)^2+m^2](k^2+m^2)}D_{\beta}
A_{\alpha}(-p,\theta)\Sigma(p,\theta)~,
\end{equation}

\noindent
while, for the second term, we found $\Gamma^{(2)}_{A\Sigma,2}=-\Gamma^{(2)}_{A\Sigma,1}$, so that the would be  $\Gamma^{(2)}_{A\Sigma}$ vanishes. Essential to this result was the choice of the order of the fields $\Sigma\Phi_a\bar\Phi_a$, instead of $\Sigma\bar\Phi_a\Phi_a$, in the action in Eq.~(\ref{actionp}). This choice, nevertheless, requires $\Sigma$ to change under gauge transformations, to keep the action gauge invariant, as we pointed out earlier. Clearly, one could choose the $\Sigma\bar\Phi\Phi$ order to keep $\Sigma$ gauge invariant, but then the Moyal phase factors in the two vertices in Eq.~(\ref{c3eq37}) would not compensate each other and, as a consequence, $\Gamma^{(2)}_{A\Sigma,1}$ and $\Gamma^{(2)}_{A\Sigma,2}$ would not cancel. In this way, a mixed term $A\Sigma$ would be generated in the effective action, making highly cumbersome the evaluation of propagators and quantum corrections.

\section{Renormalizability of the model}\label{sec4}

Let us now investigate the renormalizability of the model at the leading $1/N$ order. The power counting for the supersymmetric $CP^{(N-1)}$ model, in the fundamental representation, is the same as in the adjoint representation that was studied in~\cite{Asano:2004vq}, so we just quote the result. The superficial degree of divergence of a given supergraph is 
\begin{equation}\label{pwc}
\omega\,=\,2-\frac{1}{2}\left(E_\Phi+E_A+N_D\right)-E_\Sigma-E_c\,,
\end{equation}

\noindent
where $E_i$ is the number of external legs of the field $i$, and $N_D$ is the number of covariant derivatives acting on the external legs. We remark that, in order to have an iso-scalar contribution to the effective action, these variables are subjected to the constraints that $E_\Phi$ and the sum $E_A + N_D$ are even numbers. 

From Eq.~(\ref{pwc}) we see that, apart from vacuum diagrams, the most divergent quantum corrections to the effective action are linearly ultraviolet divergent, and these can be dangerous to the renormalizability of the model since they can generate non-integrable (linear) infrared UV/IR singularities. It is essential to secure that linear UV/IR singularities do not appear since they would invalidate the $1/N$ expansion at higher orders~\cite{Minwalla:1999px}. Some of the graphs with $\omega = 1$ have already been analyzed and shown not to generate dangerous linear divergences: the ones corresponding to the spinorial superfield effective action ($E_A=2$), calculated in Section~\ref{tpA}, and the graph with $E_\Sigma = 1$, which has been taken into account in the gap equations. The only remaining contributions with $\omega = 1 $ is the one with $(E_A,N_D)=(1,1)$, which vanishes; in fact, it is proportional to $\int d^2 \theta_1 \,D^\alpha \left( D^2+2m\right)\delta_{11}\,=\,0$. Since there is no corresponding counterterm in $S_{CT}$, the finiteness of this contribution is essential to the renormalizability of the model.

Now, we focus on graphs with logarithmic power counting. These generate integrable, and therefore harmless, UV/IR infrared singularities. However, some of these graphs are still dangerous because they can generate ultraviolet divergent contributions to the effective action which do not have corresponding counterterms in $S_{CT}$. Listing all possible graphs with $\omega=0$, one finds several such potentially dangerous corrections. The contribution with $E_\Sigma=2$ has been already analyzed and found to be finite in Section~\ref{tpsigma}, whereas the one with $(E_A,N_D,E_\Sigma)=(1,1,1)$ (the $A\Sigma$ mixing) yields a vanishing result, as shown in Section~\ref{mixing}. It still remains some harmful possibilities, namely for $E_A=4$, $E_\Phi=4$, $(E_A,E_\Sigma)=(2,1)$, and $(E_A,N_D)=(3,1)$. However, after checking that the phase factor induced by the Moyal product is planar in all these cases, one can argue that the logarithmically divergent parts of the Feynman integrals will be proportional to $\int_\epsilon d^3k \, \frac{k^{\alpha \beta}}{\left(k^2\right)^2}$, which vanishes due to the symmetric integration in the loop momentum. 

As for the remaining logarithmically divergent supergraphs, they correspond to terms present in the counterterm action and are, therefore, in principle renormalizable. An explicit verification of the renormalizability of the supersymmetric noncommutative $CP^{(N-1)}$ model would involve the calculation of the subleading corrections to several vertex functions, and one example of such a calculation is presented in the next section.

\section{Subleading corrections to the $\bar \Phi \Phi$ effective action}
\label{secphiphisub}

\begin{figure}[ht]
\begin{center}\includegraphics[width=0.6\textwidth]{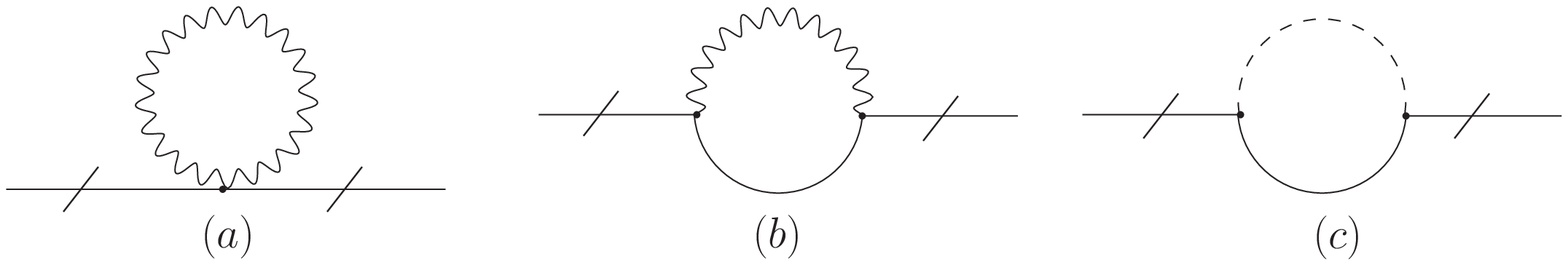}
\end{center}
\caption{Supergraphs contributing to the $1/N$ subleading contributions to the quadratic effective action of the $\Phi$ superfield.} \label{Fig4}
\end{figure}

Let us calculate in detail the subleading contributions to the quadratic effective action of the $\Phi$ superfield, which arise from the diagrams depicted in Fig.~\ref{Fig4}. 
Here the calculations become quite involved, and part of them were performed with the help of a symbolic computer program designed for superfield calculations~\cite{susymath}.
The Feynman amplitude corresponding to the graph in Fig.~(\ref{Fig4}a) is
\begin{eqnarray}\label{c3eq40}
&&\Gamma^{(2)}_{\ref{Fig4}a}  = -{\frac{i}{2}}\Big(\sum_{j=1}^{4}\int{d^3k_j\over(2\pi)^3}\Big)
\int d^2\theta~(2\pi)^3\delta^3(k_1+k_2+k_3+k_4)(2\pi)^3\delta^3(k_2+k_3)C^{\alpha\beta}\nonumber\\
&&\times \exp\Big\{-i[k_2\land(k_3+k_4)+k_3\land
  k_4]\Big\}\Big \langle
T~A_{\beta}(k_2,\theta)A_{\alpha}(k_3,\theta)\Big \rangle
\bar\Phi_a(k_1)\Phi_a(k_4)~.
\end{eqnarray}

\noindent
Integrating in $k_3$ and $k_4$, renaming $k_1=p$, $k_2=k$ and using that $\delta(\theta-\theta)=0$, we arrive at 
\begin{eqnarray}\label{c3eq41}
\Gamma^{(2)}_{\ref{Fig4}a} = {\frac{1}{N}} \int{\frac{d^3p}{(2\pi)^3}} d^2\theta\, 
\bar\Phi_a(-p,\theta) \Phi_a(p,\theta) \int{d^3k\over(2\pi)^3}\frac{-2m}{f(k)k^2(k^2+4m^2)}~.
\end{eqnarray}

\noindent
The amplitude corresponding to Fig.~\ref{Fig4}b is given by,
\begin{eqnarray}\label{c3eq43}
\Gamma^{(2)}_{\ref{Fig4}b} & = & \frac{1}{N}\int{d^3p\over(2\pi)^3}d^2\theta\,
\bar\Phi_a(-p,\theta)
\int{d^3k\over(2\pi)^3}\frac{1}{f(k)[(k+p)^2+m^2]}\nonumber\\
&\times&\Big{\{}{k^{\alpha\beta}p_{\alpha\beta}\over k^2(k^2+4m^2)}(D^2+m)
+\xi{k^{\alpha\beta}p_{\alpha\beta}\over k^4}(D^2-m)+
{2D^2\over (k^2+4m^2)}\\
&+&\xi{1\over k^2}(D^2-m)-{4m\over
  k^2(k^2+4m^2)}D^2(D^2-m)\Big{\}}\Phi_a(p,\theta)~,\nonumber
\end{eqnarray}

\noindent
and finally the contribution of Fig.~\ref{Fig4}c is
\begin{eqnarray}\label{c3eq45}
\Gamma^{(2)}_{\ref{Fig4}c} =-{1\over N}\int{d^3p\over(2\pi)^3}d^2\theta~\int{d^3k
\over(2\pi)^3}
{(D^2-m)\Phi_a (p,\theta)\bar\Phi_a (-p,\theta)\over f(k)[(k+p)^2+m^2](k^2+m^2)}~.
\end{eqnarray}

By adding the above contributions we get for the radiative corrections 
to the quadratic action in $\Phi$ the expression
 \begin{eqnarray}\label{c3eq46}
\Gamma^{(2)}_{\ref{Fig4}} & = & \frac{1}{N}\int{d^3p\over(2\pi)^3}d^2\theta\,\bar\Phi(-p,\theta)
\int_\epsilon\frac{d^3k}{(2\pi)^3}\,\frac{1}{f(k)}\Big\{
\frac{(D^2-m)^2}{k^2(k^2+4m^2)[(k+p)^2+m^2]}\nonumber\\
  & + & \frac{1}{k^2(k^2+4m^2)}
+ \xi \frac{k^2+2k \cdot p}{k^4[(k+p)^2+m^2]}\Big{\}}\left(D^2-m\right)\Phi(p,\theta)\,,
\end{eqnarray}

\noindent
which is planar and, therefore, do not generate UV/IR mixing. Despite being non-local, it still shares an overall $(D^2-m)$ factor with the piece of the classical action quadratic in $\Phi$. Separating its logarithmic divergent part,
\begin{eqnarray}\label{c3eq47}
\Gamma^{(2)}_{\ref{Fig4}}&=&\frac{1}{N}\left[(1+\xi)\int_\epsilon{d^3k\over(2\pi)^4}
\frac{1}{f(k)}\frac{1}{k^2(k^2 + m^2)}\,+\,{\rm finite terms}\right]
\nonumber\\
&\times&  \int{d^3p\over(2\pi)^3}d^2\theta \, \bar\Phi_a(-p,\theta)(D^2-m)\Phi_a(p,\theta) 
\end{eqnarray}

\noindent
we see that a wave function renormalization $\Phi\longrightarrow\sqrt{Z}\Phi$ takes care of the logarithmic UV divergence in Eq.~(\ref{c3eq47}); additionally, $\Gamma^{(2)}_{\ref{Fig4}}$ is finite in the gauge  $\xi=-1$. The important point we want to stress is that we were able to renormalize this vertex function in an explicitly supersymmetric fashion, which is not the case if one works in the Wess-Zumino gauge~\cite{Inami:2000eb}, when the bosonic and fermionic components of the superfield $\Phi$ receives different wave function renormalizations.

As for the other logarithmically divergent contributions, their calculation is much more complicated, involving graphs with one and two loops of momentum. A complete evaluation of such subleading contributions is out of the scope of this work, but in such a calculation the graphical identity represented in Fig.~\ref{figarefeva} would be essential to secure the cancellation of several UV divergences. Indeed, in the usual (commutative, nonsupersymmetric) $CP^{(N-1)}$ model, the proof of renormalizability at arbitrary order of the $1/N$ expansion heavily relies on such identity~\cite{Arefeva:1980ms}.

\section{Conclusions}\label{conc}

In this paper, we studied the noncommutative supersymmetric $CP^{(N-1)}$ model in $2+1$ spacetime dimensions, when the basic field is in the fundamental representation. We found that the model has the same phase structure as its commutative counterpart. Differently from the previous analysis in~\cite{Asano:2004vq}, in this case the model classically goes, in the $\Theta \rightarrow 0$ limit, to the usual (commutative) supersymmetric $CP^{(N-1)}$ model. At the quantum level, the UV/IR mixing generates only mild infrared singularities, so that renormalizability at the leading $1/N$ order is explicitly checked. 

The use of the superspace approach ensures a manifestly supersymmetric renormalization, which is not necessarily the case in the components fields formalism. In~\cite{Inami:2000eb}, where the ultraviolet behavior of the commutative supersymmetric $CP^{(N-1)}$ model was considered, the scalar and fermionic superpartners received different renormalizations, so that the supersymmetric invariance of the quantum theory becomes non-manifest. This problem does not appear in the superfield formalism.
 
We have also studied the first subleading correction to the effective action of the $\Phi$ field, and shown that it can be made finite by only a wave function renormalization, as it should. Such explicit calculations have not appeared in the literature so far, even in the commutative case.

\vspace{1cm}

\textbf{Acknowledgments}

This work was partially supported by the Brazilian agencies Funda\c{c}\~{a}o de Amparo 
\`{a} Pesquisa do Estado de S\~{a}o Paulo (FAPESP), Conselho 
Nacional de Desenvolvimento Cient\'{\i}fico e Tecnol\'{o}gico (CNPq),
and Coordena\c{c}\~{a}o de Aperfei\c{c}oamento de Pessoal de N\'{\i}vel Superior (CAPES). The authors thank M. Gomes for useful discussions.

\newpage

\appendix

\section{Appendix}

For the sake of clarity, we will show how the action of the noncommutative supersymmetric $CP^{(N-1)}$ model looks like when written in terms of component fields. We define the components of the spinor gauge superpotential as,
\begin{eqnarray}\label{c3eq3a}
\chi_{\alpha}(x)&=&A_{\alpha}(\theta,x)|~, \nonumber\\
B(x)&=&\frac{1}{2}D^{\alpha}A_{\alpha}(\theta,x)|~, \nonumber\\
V_{\alpha\beta}(x)&=&-\frac{i}{2}(D_{\alpha}A_{\beta}+D_{\beta}A_{\alpha})
(\theta,x)|~,\nonumber\\
\lambda_{\alpha}(x)&=&\frac{1}{2}D^{\beta}D_{\alpha}A_{\beta}(\theta,x)|~,
\end{eqnarray}

\noindent
and the components of the fields $\Phi$ and $\Sigma$ in gauge covariant way through,
\begin{eqnarray}\label{c3eq4}
\phi(x) & = & \Phi(x,\theta)|~,\nonumber\\
\psi_{\alpha}(x) & = & \nabla_{\alpha}\Phi(x,\theta)|~,\\
F(x)  & = & \nabla^2\Phi(x,\theta)|~,\nonumber
\end{eqnarray}

\noindent
and
\begin{eqnarray}\label{c3eq5}
\kappa(x) & = & \Sigma(x,\theta)|~,\nonumber\\
\zeta_{\alpha}(x) & = & \nabla_{\alpha}\Sigma(x,\theta)|~,\\
\sigma(x) & = & \nabla^2\Sigma(x,\theta)|~,\nonumber
\end{eqnarray}

\noindent
where the instruction $|$ means to take $\theta=0$ after doing the 
derivatives. Using these definitions, the action in Eq.(\ref{c3eq2}) can be cast as
\begin{eqnarray}\label{c3eq6}
S & = & \int d^3x\Big{\{}\bar{F}_aF_a+\bar{\psi_a}^{\alpha}(i{\partial_{\alpha}}^{\beta}+{V_{\alpha}}^{\beta}){\psi_a}_{\beta}
\nonumber\\
& + &(i\bar{\psi_a}^{\alpha}\lambda_{\alpha}\phi_a+~c.h.)+
(\partial_{\alpha\beta}\bar\phi_a+i\bar\phi_aV_{\alpha\beta})(\partial_{\alpha\beta}\phi_a-iV_{\alpha\beta}\phi_a)
\nonumber\\
  & - & \sigma\left(\phi_a\bar\phi_a-{\frac{N}{g_0}}\right) - 
\zeta^{\alpha}({\psi_a}_{\alpha}\bar\phi_a+\phi_a \bar{\psi}_{a\,\alpha}) \nonumber\\
&-&\kappa(F_a\bar\phi_a+\phi_a\bar{F_a}+\psi^{\alpha}_a \bar{\psi}_{ a\,\alpha})
\Big{\}}~.
\end{eqnarray}

\noindent
As in the main text of this paper, the Moyal product is not written explicitly in this and in the remaining formulas. The auxiliary fields $F$ and $\bar F$ can be eliminated by means of their equations of motion and, in this way, Eq.~(\ref{c3eq6}) is reduced to
\begin{eqnarray}\label{c3eq17b}
&&S  =  \int d^3x\Big{\{}\bar{\psi_a}^{\alpha}(i{\partial_{\alpha}}^{\beta}
+{V_{\alpha}}^{\beta}){\psi_a}_{\beta}
+(\partial_{\alpha\beta}\bar\phi_a+i\bar\phi_aV_{\alpha\beta})(\partial_{\alpha\beta}\phi_a-iV_{\alpha\beta}\phi_a)
\nonumber\\
&&-\sigma\left(\phi_a\bar\phi_a-{N\over g_0}\right)
 -  \bar\phi_a(\zeta^{\alpha}+i\lambda^{\alpha}){\psi_a}_{\alpha}
-\bar{\psi_a}_{\alpha}(\zeta^{\alpha}-i\lambda^{\alpha})\phi_a \nonumber\\
&& -\kappa^2\phi_a\bar\phi_a-
\kappa\psi^{\alpha}_a\bar{\psi_a}_{\alpha}\Big{\}}~.
\end{eqnarray}

Finally, by writting the bi-spinors in terms of the more usual 3-vectors 
through $V_{\alpha\beta}=1/2(\gamma^{\mu})_{\alpha\beta}A_{\mu}$ and
$\partial_{\alpha\beta}=1/2(\gamma^{\mu})_{\alpha\beta}\partial_{\mu}$,
where $\gamma^{\mu}$ are the three $2\times2$ Dirac matrices in $2+1$ 
dimensions, we arrive at
\begin{eqnarray}\label{c3eq17c}
&&S  =  \int d^3x\Big{\{}\bar\phi_a \Box \phi_a-\sigma\left(\phi_a\bar\phi_a
-\frac{N}{g_0}\right)
-i\left(\bar\phi_a \, \overset{\leftrightarrow}{\partial^{\mu}} \, \phi_a\right)A_{\mu}
+\bar\phi_aA^2\phi_a \nonumber\\ 
&& + i \bar\psi_a \gamma^\mu \left(\partial_\mu-iA_{\mu}\right)\psi_a
-\bar\phi_a(\zeta+i\lambda)\psi_a-\bar{\psi_a}(\zeta-i\lambda)\phi_a \nonumber\\
&& -\kappa^2\phi_a\bar\phi_a-\kappa\psi_a\bar\psi_a\Big{\}}~.
\end{eqnarray}

\noindent
This expression shows more explicitly what is the theory we are 
working with, when dealing with the more compact superfield notation. One can realize that the first line corresponds to the bosonic model of Eq.~(\ref{c3eq1}), extended to the noncommutative spacetime,
but we now have the additional fermionic degrees of freedom, necessary for supersymmetry, and a new constraint imposed by the combination $\zeta+i\lambda$, and $\kappa$ is a composite field classically given by $\kappa = - \psi_a \bar \psi_a / 2 \phi_a \bar\phi_a $. Elimination of $\kappa$ in Eq.~(\ref{c3eq17c}) yields a four-fermion self-interaction, typical of the supersymmetric extension of the $CP^{(N-1)}$ model~\cite{Abdalla:1990qf}.

\end{document}